\documentclass[prb,aps,twocolumn,showpacs]{revtex4}
\usepackage{epsfig}
\begin{document}

\title{X-ray diffraction measurements of the c-axis
Debye-Waller factors of YBa$_2$Cu$_3$O$_7$ and
HgBa$_2$CaCu$_2$O$_6$}

\author{C. Kim$^{1,2}$, A. Mehta$^1$, D.L. Feng$^{3,4}$, K.M. Shen$^3$,
N.P. Armitage$^3$, K. Char$^5$, S.H. Moon$^6$, Y.Y. Xie$^7$, and
J. Wu$^7$}

\affiliation{$^1$Stanford Synchrotron Radiation Laboratory,
Stanford, California 94309}

\affiliation{$^2$Institute of Physics and Applied Physics, Yonsei
University, Seoul, Korea}

\affiliation{$^3$Department of Physics and Applied Physics,
Stanford University, California 94305}

\affiliation{$^4$Department of Physics, Fudan University, Shanghai
200433, People's Republic of China}

\affiliation{$^5$School of Physics and Center for Strongly
Correlated Materials Research, Seoul National University, Seoul
151-742, Korea}

\affiliation{$^6$LG Electronics Institute of Technology, Seoul
137-724, Korea}

\affiliation{$^7$Department of Physics and Astronomy, University
of Kansas, Lawrence, Kansas 66045}
\date{\today}
\begin{abstract}
We report the first application of x-rays to the measurement of
the temperature dependent Bragg peak intensities to obtain
Debye-Waller factors on high-temperature superconductors.
Intensities of $(0,0,l)$ peaks of YBa$_2$Cu$_3$O$_7$ and
HgBa$_2$CaCu$_2$O$_6$ thin films are measured to obtain the c-axis
Debye-Waller factors. While lattice constant and some Debye-Waller
factor measurements on high T$_c$ superconductors show anomalies
at the transition temperature, our measurements by x-ray
diffraction show a smooth transition of the c-axis Debye-Waller
factors through T$_c$. This suggests that the dynamic
displacements of the heavy elements along the c-axis direction in
these compounds do not have anomalies at T$_c$. This method in
combination with measurements by other techniques will give more
details concerning dynamics of the lattice. 
\end{abstract}
\maketitle

The majority of efforts to understand the mechanism of the high
T$_c$ superconductivity (HTSC) have been in terms of pure
electronic effects, primarily because of the conjecture that
electron-phonon coupling alone cannot give such a high T$_c$.
However, there have been persistent efforts to study the role of
the lattice in
HTSC\cite{Nazarenko,Sakai,You,Asahi,Kadowaki,Braden,Pasler}. In
addition to the fact that the structural information provides
important information about the materials, the lattice can still
play an important role as it affects both the hopping ($t$) and
magnetic exchange ($J$) energies of these materials, the two most
important parameters in the physics of these correlated systems.
Indeed, there have been theoretical efforts to study the role of
lattice vibration in the cuprates\cite{Nazarenko,Sakai}.
Therefore, experimental investigation of the temperature dependent
lattice vibration in these materials, especially the critical
behavior around T$_c$, is essential and can provide vital
information concerning the low energy degrees of freedom.

Experimental evidence that shows clear correlation between the
critical behaviors in the electronic and crystal structures are
anomalies in lattice constants at T$_c$. These have been observed
in various HTSC materials by x-ray\cite{You,Asahi} and more
accurately by capacitance dilatometry
methods\cite{Kadowaki,Braden,Pasler}. They have shown that the
lattice constant generally decreases at a much faster rate at
T$_c$ than at other temperatures as the temperature decreases.
Subsequently, the thermal expansion coefficient $\alpha$ shows an
anomaly that is of similar shape to the heat capacity anomalies
observed at T$_c$. This enhanced decrease in the lattice constant
has been related to the specific heat capacity through the
thermodynamic Ehrenfest relations\cite{Pasler}. Yet the
microscopic reason behind such a lattice anomaly is not well
understood. One natural question is if it is related to the
dynamic properties of phonons. It will therefore be interesting to
measure the Debye-Waller factors (DWF) of HTSC materials.

In complex materials, the DWFs are usually measured by extended
x-ray absorption fine structure (EXAFS). In the HTSCs there have
been efforts to measure the temperature dependent mean-square
displacement $\sigma ^2(T)$ (which is inversely related to the
DWF) by EXAFS\cite{Yang,Bridges,Lanzara}, with essentially only Hg
based samples showing anomalies at T$_c$\cite{Lanzara}. EXAFS
measures local coordinates at a certain atom and therefore has an
important advantage of being element specific. It, however,
measures only relative atomic motions (for example, O relative to
Cu or optical phonons in the Cu-O plane) and is insensitive to
collective motions of the atoms, for instance, the low frequency
acoustic phonons. To see the effect in the phonon population, the
necessary condition is to have ample phonons in the first place.
This may be difficult for optical phonons as their energy scale is
usually very high compared to the experimental temperature scales
(which are set by $T_c$). This may be related to the fact that
EXAFS results largely show no anomalies. In contrast, even though
they have their own disadvantages which will be discussed later,
diffraction techniques in principle measure the integrated phonon
effects\cite{Kittel}. Here we report novel application of x-ray
diffraction that can in principle measure DWFs. This is to our
knowledge the first application of this method by x-rays to
complex materials like HTSCs. We have measured the temperature
dependence of the Bragg peak intensities, a quantity that can be
related to the dynamic lattice motion. The results are discussed
along with various aspects of x-ray diffraction and possible
future experiments.

The experiments were performed at beamline 2-1 of the Stanford
Synchrotron Radiation Laboratory (SSRL) which is equipped with a
2-axis Huber diffractometer. The diffractometer has an open cycle
He cryostat. $h\nu =8.8$ keV x-rays, just below the Cu K edge,
were used to reduce the background from the fluorescent light. For
the reasons explained below, intensities of the $(0, 0, l)$ peaks
of YBa$_2$Cu$_3$O$_7$ (YBCO) and HgBa$_2$CaCu$_2$O$_6$ (Hg1212)
thin film samples grown on SrTiO$_3$ substrates were taken. These
materials were grown according to the procedures described
elsewhere\cite{Cole,Wu}. The YBCO film has T$_c$ of 90.5K with the
$\Delta$T$_c$ less than 1K. The Hg1212 sample has the
$\Delta$T$_c$ less than 1K locally but shows distribution of
T$_c$'s from 123K at the center to 118K at the edge.

In order to obtain highly reliable intensities, it is essential to
reduce the errors from the beam and sample instabilities and
detector non-uniformity. The beam drifts were monitored in the
I$_0$ section and corrected for before each measurement. The
biggest effect came from the motion of the cryo-head during the
temperature cycle. To reduce such effects, a special manipulator
has been designed as shown in Figure 1. The sample holder is in
weak thermal contact with the cryo-head and has its own
temperature sensor and heater. The thermal contact is weak enough
so that the temperature of the sample holder can be varied between
10 K and 200 K with the temperature of the cryo-head maintained at
10 K. This removed most of the sample motion and greatly improved
the reproducibility of the data.

The special manipulator did not completely remove the sample
motion with respect to the beam during temperature cycling. If the
sample was completely homogeneous then the relative motion of the
sample will have no effect on the measurement. In our experience,
the only high T$_c$ samples which are homogeneous to a highly
collimated and monochromatic synchrotron beam are fine grain
powders. However, in a powder sample the high index peaks, the
peaks more sensitive to the DWF (as DWF $\sim l^2$), were too weak
to measure with sufficient statistical accuracy in a reasonable
amount of time. Even though the intensities of the high index
Bragg peaks were acceptable in a single crystal we discovered,
after looking at high T$_c$ single crystals from several different
sources, that these samples were really a bundle of several low
angle grain boundary crystallites.  Even a small motion of such a
sample with respect to the incident beam changed the fraction of
the diffracting domain in the beam, which resulted in a
significant change in the diffracted beam intensity.  Thin films
of sufficiently high quality and lateral homogeneity for
synchrotron measurements were available but they were all $(0, 0,
l)$ orientation\cite{Side}. Hence, as a compromise between high
intensity and sufficiently homogeneous samples, we performed our
measurements on the $(0, 0, l)$ peaks of thin film samples.

\begin{figure}[t]
\centering \leavevmode \epsfxsize=7.5cm \epsfbox{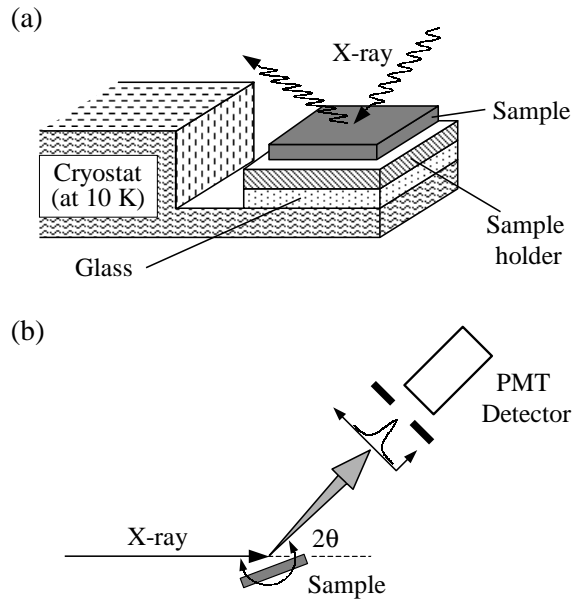}
\vspace{.2cm} \caption{(a) The schematic of the cryostat designed
for the experiments. The sample holder is in weak thermal contact
with the cryo-head with its own temperature sensor and heater. Due
to the weak thermal contact, the temperature of the sample holder
can be varied between 10 K and 200 K while the temperature of the
cryo-head fixed at 10 K. (b) Illustration of the experimental
configuration. The detector is fixed at the 2$\theta$ position
while $\theta$ is scanned. The detector slit is wide enough to
accept the whole Bragg peak as illustrated unlike the conventional
setup. The 2$\theta$ position is adjusted at each temperature so
that the same part of the detector is used.}\label{fig1}
\end{figure}

In addition to the motion of the cryostat upon the temperature
change, other factors such as detector non-uniformity contributed
to the irreproducibility of the data. Since the effect we were
looking for was very small, the accuracy/reproducibility of the
measurements had to be better than 0.5$\%$. Therefore, care had to
be taken to remove any factor that affects the data quality. The
following somewhat unconventional procedure was performed to
obtain reliable data. First, the detector slit was set wide enough
($\Delta 2\theta \sim$ 4 degrees) to accept all the diffracted
x-ray within a certain Bragg peak (Fig. 1). This is to minimize
the detector non-uniformity effect as use of angle limiting
devices such as Soller slits resulted in very unreliable data.
However, some degree of non-uniformity still existed and it was
necessary to make it sure that the same part of the detector is
used for each scan. To ensure this, we scanned 2$\theta$ at each
temperature to locate the detector so that the centroid of the
diffraction peak is at the center of the detector. In this way,
the detector is essentially tracking the temperature dependent
motion (in 2$\theta$) of the diffraction peak. As the last step,
the $\theta$ scans (rocking curves) were taken to measure the
diffraction peak intensity. The above procedure produced the most
reliable data.

Fig. 2 shows $\theta$ scans as well as temperature dependent
intensity plots measured as described above. Panel (a) shows
normalized $\theta$ scans of the (0,0,13) peak from YBCO. The peak
position in $\theta$ decreases as temperature increases, implying
the increase of the c-axis lattice constant as expected. In
addition to the decrease in the $\theta$ position, we note that
the Bragg peak intensity decreases as the temperature increases
(increased constant background due to incoherent scattering was
also observed but subtracted in the plot and analysis). No
appreciable change in the diffraction line shape, that is, no $q$
dependence is observed and thus the role of thermal diffuse
scattering is not considered in the following discussions. Since
the detector slit was wide open (in $2\theta$) to accept all the
x-rays of the peak, each point on the $\theta$ scan represents
Bragg peak intensity i.e., integration of a $2\theta$ scan at the
given $\theta$ value. Therefore, the integration of the area
represents the sum of the intensities from all the grains of the
sample. It is apparent from the figures that the peak intensity
decreases as the temperature increases.

\begin{figure}[t]
\centering \leavevmode \epsfxsize=8.5cm \epsfbox{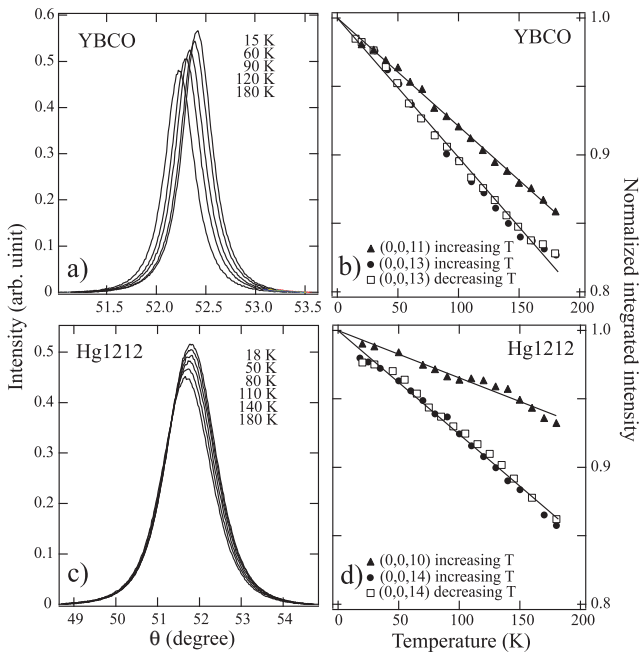}
\vspace{.2cm} \caption{(a)Series of $\theta$ scans of (0,0,13)
peak from YBCO at different temperatures. The changes in total
intensity as well as the peak position are seen. (b)Temperature
vs. integrated intensity plot of the data. The data were taken
with the temperature cycled and normalized at T=0. Also shown is
the data for (0,0,11) peak. (c)Scans of (0,0,14) from Hg1212
samples. (d)Temperature vs. intensity plot for (0,0,14) and
(0,0,10) peaks. }\label{fig2}
\end{figure}

To quantify the temperature dependent peak intensities, the peaks
shown in Fig. 2a are integrated over the whole angular range. The
results are normalized to a common linear extrapolated T=0 value
and plotted against the temperature in panel (b). For monatomic
systems, the intensity of a (0,0,$l$) Bragg peak is expressed as
$I=I_0\exp(-\frac{1}{2}\langle z^2\rangle l^2)$ where the exponent
is the DWF and $\langle z^2\rangle$ is the thermally averaged mean
square motion of the atoms in z direction\cite{DWF}. For a
classical Harmonic oscillator, $\langle z^2\rangle\propto T$ and
the intensity thus becomes $I=I_0\exp(-\lambda l^2 T)\propto
I_0(1-\lambda l^2 T)$ where $\lambda$ is a constant. Therefore,
one would expect linear temperature dependence of the DWFs for
T$\ll 1/\lambda l^2$. The intensities in the figure show linear
temperature dependence and no anomaly is found within the
experimental error\cite{ErrorBar}. Also shown in the panel is the
temperature dependent intensity of the (0,0,11) peak from YBCO.
The ratio of the DWFs of the (0,0,13) and (0,0,11) peaks is
$1.28\pm 0.05$ which is somewhat close to the expected value of
$13^2/11^2=1.4$ for monatomic systems with pure thermal phonon
effects. The measured temperature dependent intensities suggest
normal behavior of the c-axis DWF without any anomaly. Panels (c)
and (d) show data from Hg1212 samples. Other than the larger
rocking width, the behavior is more or less the same with YBCO
case, showing no sign of anomaly at $T_c$ within the experimental
error. The deviation of (0,0,10) data from the linear behavior is
related to the fact that, unlike other peaks, it was measured
along with substrate peaks hence producing larger experimental
errors, and is assumed to be extrinsic effect.

Interpretation of the above results, however, is not as easy as
the case for monatomic systems such as Si. The expression of a
Bragg peak intensity contains in general exponential functions of
the displacement parameters of the atoms in the unit cell. For a
monatomic case, it could be reduced to a single exponential
function of the displacement parameter (assuming the displacement
parameters for the atoms in the unit cell are the same) and the
DWF can be interpreted with a relative ease. The form factors for
polyatomic compounds however are different and the Bragg peak
intensity is expressed by a combination of exponential functions.
It is therefore very hard to extract the information on the
dynamic displacements from DWFs. The other aspect is that the
observed temperature dependence does not solely come from the
dynamic displacements. Not only the phonons but also the changes
in the static atomic positions within the unit cell affect the
intensities.

In spite of the difficulties, we can still extract useful
information from the results. Even though the Bragg peak intensity
expression contains multiple exponential terms, it shows linear
temperature dependence if the displacements are purely due to
thermal vibrations. Therefore, the fact that experimental data
show linear temperature dependence strongly suggests that the
major contributor to the DWF is the dynamic displacements as there
is no intrinsic reason for such temperature dependence from the
static displacements. Indeed, detailed neutron diffraction
experiments on YBCO show that the mean square displacements are
mostly dynamic\cite{Schweiss}. We also note that heavy elements
contribute more to the diffraction intensities in our x-ray
measurements due to their greater high Z sensitivity. To see the
different contributions, we list in Table 1 calculated fractions
of the planar oxygen and Cu-O planar contributions to the total
diffraction peak intensities of YBCO and Hg1212 using the
published structural information\cite{Jorgensen,Morosin}. Note
that the fractions do not add up to make the total due to the
interference but roughly show the insignificant Cu-O plane
contribution to the total intensity. It is thus conceivable that
the phonon anomalies in the Cu-O plane measured by
EXAFS\cite{Lanzara} may be buried under the contributions from the
heavy elements. Based on these, it is reasonable to conclude that
at least the heavy elements do not show anomaly at $T_c$ in their
c-axis dynamic properties.

On the more fundamental side, the anomalies in the c-axis lattice
constants are generally relatively small compared to those along
a- or b-axes\cite{Asahi,Braden,Pasler}. Therefore, it is possible
that the c-axis phonon population anomaly may well be very small
and undetectable. Looking at a- and b-axis Debye-Waller factors is
therefore essential. This may explain the absence of the anomaly
in the Hg1212 data while measurement by EXAFS on powder samples of
similar compound HgBa$_2$CuO$_4$ shows anomaly in the mean square
dispacements\cite{Lanzara}. Use of Rietveld refinement with powder
diffraction at a brighter x-ray source could give an answer to
this. Better yet, considering all the factors discussed above,
similar experiments by neutron diffraction may resolve most of the
problems; the typical grain sizes of HTSC crystals and sample
motion upon temperature change are insignificant compared to the
neutron beamsize. This will allow us to investigate a- and b-axis
DWFs by a diffraction technique. In addition, its sensitivity to
low Z atoms will allow more accurate estimation of the oxygen
contribution to the DWFs. There indeed is a strong indication that
temperature dependent (1,2,12) neutron Bragg peak intensities of
YBCO crystals have anomalies at the T$_c$\cite{Schweiss}.

In conclusion, temperature dependent c-axis DWFs of YBCO and
Hg1212 thin films are obtained by measuring the temperature
dependent $(0,0,l)$ x-ray Bragg peak intensities. The measured
DWFs show linear temperature dependence within the experimental
errors, showing no anomalous effects near T$_c$. Considering the
fact that the x-ray diffraction intensity is much more sensitive
to the heavy elements, the results suggest that there is no
anomalous lattice dynamics of heavy elements along the c-axis.
Further studies of a- and b-axis temperature dependent DWFs by
neutron diffraction in combination with temperature dependent
structural studies by EXAFS may shed more light on the temperature
dependent lattice dynamics, i.e., role of the phonons in HTSCs.

\begin{table}
  \begin{center}
  \item[]\begin{tabular}{|c|c|c|c|}
  \hline
  Sample & $F_c^2$(total) & $F_c^2$(Cu-O plane) & $F_c^2$(planar O) \\
  \hline
  YBCO & 50.0 & 2.58 & 1.00 \\
  Hg1212 & $1386$ & $1.90$ & $1.00$ \\ \hline
\end{tabular}
  \caption{Calculated $F_c^2$'s for YBCO (0,0,13)and Hg1212 (0,0,14)
    normalized to plane O values. Cu-O plane and planar oxygen
    contributions have been calculated with the occupancies of the
    other atoms set to zero.}
  \end{center}
\end{table}

We would like to thank F. Bridges and J. Arthur for helpful
discussions, and H. Shin for statistical analysis. SSRL is
operated by the DOE office of Basic Energy Research, Division of
Chemical Sciences. The office's division of Material Science
provided support for this research. This work is supported (in
part) by the Korean Science and Engineering Foundation (KOSEF)
through Center for Strongly Correlated Materials Research (CSCMR)
at Seoul National University.


\begin{thebibliography}{15}
\bibitem{Nazarenko}A. Nazarenko, and E. Dagotto, Phys. Rev. B {\bf 53}, 2987 (1996).
\bibitem{Sakai}T. Sakai, D. Poilblanc, and D. J. Scalapino, Phys.Rev. B {\bf 55}, 8445 (1997).
\bibitem{You}H. You, U. Welp, and Y. Fang, Phys. Rev. B {\bf 43}, 3660 (1991).
\bibitem{Asahi}T. Asahi, H. Suzuki, M. Nakamura, H. Takano, and J. Kobayashi,
Phys. Rev. B {\bf 55} 9125 (1997).
\bibitem{Kadowaki}K. Kadowaki, F. E. Kayzel, and J. J. Franse, Physica C {\bf 153-155}, 1028 (1988).
\bibitem{Braden}M. Braden, O. Hoffels, W. Schnelle, B. B\"{u}chner, G. Heger, B. Hennion,
I. Tanaka, and H. Kojima, Phys. Rev. B {\bf 47}, 12288 (1993).
\bibitem{Pasler}V. Pasler, P. Schweiss, C. Meingast, B. Obst, H. W\"{u}hl, A. I.
Rykov, and S. Tajima, Phys. Rev. Lett. {\bf 81}, 1094 (1998).
\bibitem{Yang}C. Y. Yang, S. M. Heald, J. M. Tranquada, A. R. Moodenbaugh, and
Y. Xu, Phys. Rev. B, {\bf 38}, 6568 (1988).
\bibitem{Bridges}C. H. Booth, F. Bridges, J. B. Boyce, T. Claeson, B. M. Lairson,
R. Liang, and D. A. Bonn, Phys. Rev. B {\bf 54}, 9542 (1996).
\bibitem{Lanzara}A. Lanzara, N.L. Saini, A. Bianconi, F. Duc, and P. Bordet,
Phys. Rev. B, {\bf 59}, 3851 (1999).
\bibitem{Kittel}C. Kittel, {\it Introduction to Solid State Physics},
         John Wiley \& Sons, Inc., New York, 1986.
\bibitem{Cole}B. F. Cole, G.-C. Liang, N. Newman, K. Char, G. Zaharchuk, and
J. S. Martens, Appl. Phys. Lett. {\bf 61} 1727 (1992).
\bibitem{Wu}J. Z. Wu, S.L. Yan, and Y.Y. Xie, Appl. Phys. Lett. {\bf 74} 1469
(1999).
\bibitem{Side} a- or b-axis YBCO thin films do exist. However,
these films are grown on LaSrGaO$_4$ substrates which have a cubic
structure. The result is that the substarte peaks, which are much
stronger, overlap with the peaks from the film, making those film
sample useless for our experiments.
\bibitem{DWF}This formula holds true only for monatomic systems.
For polyatomic compounds, it is more complicated. See the
discussion that follows later in the text.
\bibitem{ErrorBar}The fluctuations in Fig 2b and 2d are not from
the statistical shot noise in the counts but are mostly due to the
extrinsic effects such as the thermal motion of the sample as
discussed in the text. The size of the error bar based on our
experience is 0.5$\%$ or less. All of the root mean square errors
of the data are less than 0.5$\%$.
\bibitem{Jorgensen}J. D. Jorgensen, B. W. Veal, A. P. Paulikas, L. J. Nowicki,
G. W. Crabtree, H. Claus, and W. K. Kwok , Phys. Rev. B, {\bf 41}, 1863 (1990).
\bibitem{Morosin}For Hg1212, published atomic positions of TlBa$_2$CaCu$_2$O$_7$ which
is isostructural with Hg1212 were used. For TlBa$_2$CaCu$_2$O$_7$
structure, see B. Morosin, D. S. Ginley, P. F. Hlava, M. J. Carr,
R. J. Baughman, J. E. Schirber, E. L. Venturini, and J. F. Kwak,
Physica C {\bf 152}, 413 (1998).
\bibitem{Schweiss}P. Schweiss, W. Reichardt, M. Braden, G. Collin, G. Heger,
H. Claus, and A. Erb, Phys. Rev. B, {\bf 49}, 1387 (1994).
\end{thebibliography}
\end{document}